\def\year{2022}\relax
\documentclass[letterpaper]{article} 
\usepackage{aaai22}  
\usepackage{times}  
\usepackage{helvet}  
\usepackage{courier}  
\usepackage[hyphens]{url}  
\usepackage{graphicx} 
\usepackage{natbib}  
\usepackage{caption} 
\DeclareCaptionStyle{ruled}{labelfont=normalfont,labelsep=colon,strut=off} 
\usepackage{algorithm}
\usepackage{algorithmic}
\usepackage{booktabs}
 \usepackage[table,xcdraw]{xcolor}
\usepackage{colortbl}
\usepackage{newfloat}
\usepackage{listings}
\floatstyle{ruled}
\newfloat{listing}{tb}{lst}{}
\floatname{listing}{Listing}
\title{An Investigation of Experiences Engaging the Margins in Data-Centric Innovation}

\author{
    Gabriella Thompson,\textsuperscript{\rm 1}
    Ebtesam Al Haque,\textsuperscript{\rm 2}
    Paulette Blanc,\textsuperscript{\rm 3}\\
    Meme Styles,\textsuperscript{\rm 3}
    Denae Ford,\textsuperscript{\rm 4}
    Angela D. R. Smith,\textsuperscript{\rm 1}
    Brittany Johnson\textsuperscript{\rm 2}
}
\affiliations{
    \textsuperscript{\rm 1} University of Texas at Austin\\

    gabriella.thompson@utexas.edu, angela.smith@ischool.utexas.edu

    \textsuperscript{\rm 2} George Mason University\\
    ehaque4@gmu.edu, johnsonb@gmu.edu

    \textsuperscript{\rm 3} Measure \\
    paulette@wemeasure.org, meme@wemeasure.org

    \textsuperscript{\rm 4} Microsoft Research\\
    denae@microsoft.com
}

\begin{document}

\maketitle

\begin{abstract}

Data-centric technologies provide exciting opportunities, but recent research has shown how lack of representation in datasets, often as a result of systemic inequities and socioeconomic disparities, can produce inequitable outcomes that can exclude or harm certain demographics.
In this paper, we discuss preliminary insights from an ongoing effort aimed at better understanding barriers to equitable data-centric innovation.
We report findings from a survey of 261 technologists and researchers who use data in their work regarding their experiences seeking adequate, representative datasets.
Our findings suggest that age and identity play a significant role in the seeking and selection of representative datasets, warranting further investigation into these aspects of data-centric research and development.


\end{abstract}

\section{Introduction \& Background}

Data is at the center of modern research and development, often
providing generalized insights into people and phenomena~\cite{wu2013data}.
While there is potential in the power of data to offer personalized benefits to broader society, this power is diminished when there are gaps in the data being used. 
Data-centric innovations in technology continue to demonstrate these gaps through their failure to equally support all users~\cite{miller2020matter}.
Several studies on algorithmic fairness suggest representation bias in training data is a contributing factor to the high error rates for users with historically marginalized identities~\cite{buolamwini2018gender, markl2022language, baack2024critical, chen2018my, lin2020identifying, asudeh2019assessing}. 
Attempts to diversify datasets often promote inclusive design methods, such as expanding the possibilities for self-identification on demographic forms~\cite{bivens2016baking, slade2021sex}, or propose novel methods for diverse data collection~\cite{stasaski2020diverse, lin2020identifying}, but there is a gap in scholarship on the experiences of seeking diverse data from the technologists’ perspective. 
Uncovering the common successes and limitations they encounter may illuminate the barriers to equitable data-centric research and development. 
In this paper, we report on preliminary findings from a survey of technologists regarding their experiences engaging in data-centric work.

 

\section{Methodology}
To understand our respondents’ experiences seeking data, we developed three research questions that guided our survey design and analysis:  \textit{What are the factors that impact technologists' decision to use a dataset?} (\textbf{RQ1}), \textit{What are the challenges and barriers to finding diverse or representative datasets?} (\textbf{RQ2}), and \textit{What methods do technologists use to find relevant and representative datasets? } (\textbf{RQ3}).\\




\noindent\textbf{Survey Design \& Dissemination.}
We developed our survey as part of a larger effort to understand the barriers of engaging marginalized groups in data-centric computing research and development. 
The survey is designed for two audiences: technology users who identify as Black, Indigenous, or People of Color (BIPOC), and technologists and researchers who use data in their work. 
Our survey consists of four sections: \textit{Contributing Data}, \textit{Seeking Data}, \textit{Collecting Data - Research}, and\textit{ Demographics}. 
In this paper, we focus on the technologists and researchers who completed the Seeking Data section.
The questions in the Seeking Data section centered on methods used to search for data, how they make decisions about the data they will use, and their experiences finding adequate data. 
We also included an attention question for all participants to ensure they were giving the survey due consideration. 
Our full survey is publicly available for reuse and replication.~\footnote{\url{https://inspired-gmu.github.io/engaging-margins/#goals}}
We administered our survey using Qualtrics.~\footnote{\url{https://www.qualtrics.com/}} 
To recruit respondents, we advertised on social media (e.g., X, formerly known as Twitter, LinkedIn) and in our professional networks. From these efforts, we acquired over 900 survey responses. Because we administered the survey online~\cite{griffin2021ensuring}, we proceeded to clean the data of any invalid responses.\\


\noindent\textbf{Data Preparation \& Cleaning.}
To ensure validity of our data, we first removed incomplete responses and those that took less than 3 minutes to finish. We then filtered out duplicate responses (based on email) and irrelevant open ended responses. 
To prepare our data for analysis, we combined the age responses into two new categories: \textit{Under 35} (135 responses) and \textit{35 and Over} (124 responses).
This decision was informed by insights from the StackOverflow Developer Survey~\footnote{\url{https://survey.stackoverflow.co/2023/}}, which found that 43\% of professional developers are within the ages of 24--35. We also removed respondents out of scope for our analysis.\\

\noindent\textbf{Respondents.}
From our cleaned dataset, we found most technologists work in industry, either in a technical (196) or research role (27), followed by academia (27).
We had a handful of respondents from other occupations, such as healthcare (4) and childcare (5).
Of the 261 respondents in our sample, 135 are between the ages of 18--34 years old, while 124 are between 35-84. The majority of respondents in the 35--84 age group are under 64 years old (94\%).
The majority of our respondents (249) identified as a Person of Color (POC). 
A handful of respondents seek data for technology development alone (37), but the majority in our sample seek data for research (118) or for both. \\



\noindent\textbf{Data Analysis.}
To answer \textbf{RQ1} and \textbf{RQ2}, we focused on three independent variables (Data Use purposes, Age, and POC Declaration) across all of our statistical tests.
To determine the factors that impact the decision to use a dataset (\textbf{RQ1}), we identified five factors (Cost, Diversity, Trust in the Source, and Amount of Data in the Source) and compared responses to relevant survey questions against the independent variables to determine association. 
We identified the challenges and barriers to finding diverse data (\textbf{RQ2}) by analyzing their responses to the questions regarding how often they do not find adequate data, how easily they find trustworthy resources for data, and factors that impact their ability to find diverse datasets.
Lastly, we analyzed their methods for finding relevant and representative datasets (\textbf{RQ3}) by comparing responses to questions regarding their ability to find adequate data
and trustworthy data 
to their methods for finding data (where they start their search and where they have the most success). 

We used Python and the Pandas package \cite{zenodo2024pandas} for the majority of our data analysis. We conducted Chi-square tests of independence to determine association for comparisons that involved two categorical variables. Building on methods from prior work~\cite{sharpe2015your}, we conducted post-hoc testing for the Chi-Square tests with contingency tables larger than 2x2 to determine which relationships were driving the significance. 
We then used the Researchpy package \cite{bryant2018research} and information from Peter Statistics \cite{statisticnominal} to calculate Cramer’s V for each significant Chi-Square test to determine the strength of the association. 

For the tests that did not satisfy the requirements of the Chi-Square test, 
we conducted a two-tailed Fisher's Exact Test using the Scipy package \cite{2020SciPy-NMeth} to determine significance. 
If the contingency table was larger than 2x2, we utilized the Fisher-Freeman-Halton Exact Test. 
Lastly, we performed a Mann-Whitney U test to compare against ordinal dependent variables. 
For tests that compared more than two ordinal dependent variables, we used a Kruskal-Wallis test to determine if there was a significant difference in means. \\

\section{Findings}
In this section, we describe the findings from our survey relative to our three research questions.

\subsection{Dataset Decision Factors (RQ1)}
We identified cost, diversity, dataset structure, trust, and size of the dataset as influential factors in the decision to use a dataset. We compared each factor to three independent variables: Data Use Purposes, Age Group, and POC. Here, we discuss the significant results from the statistical tests we conducted.

\subsubsection{Cost.} 
We found a strong correlation between cost and data use purposes ($p=0.0087$).
Our post-hoc testing revealed a significant relationship between collecting data for technology development and not paying for data ($p=0.015$). We affirmed the strength of this association by calculating the Cramer’s V, which indicated that there was a small but significant association ($V=0.1906$) between the two variables. We did not find any significant relationships from the remainder of our tests.

\subsubsection{Diversity.} 
Our analyses using the Mann-Whitney U Test identified an association between the importance of data diversity and age group ($U=10969$, $p=1.2651e-6$). 
We assigned a rank for the four answer choices in the relevant question (Not important at all = 1; Not very important = 2, Kind of important = 3, and Very important = 4). 
We conducted a one-sided Mann-Whitney U test to infer the direction of the significance and found a significant p-value for the 'Under 35' age group  ($p=6.325e-7$). 
The mean for the 'Under 35' age group was 3.759, and the mean for the '35 and over' was 3.403. 
These results indicate that respondents under 35 may be more likely to rate diversity as more important to their work.
We did not find any significant results from the remainder of our tests.

\subsubsection{Trust.} 
This Chi-square test indicated a significant relationship between respondents who felt trust contributed the most to their decision to use a dataset and their age ($p=1.4136e-05$)
We compared the expected and observed counts of the Chi-square test and found that respondents over 35 relied more on trust as a factor in their decision than older technologists.
The strength of this association was small but significant according to the Cramer’s V ($V=0.2767$). 
We also found a significant relationship between identifying as a POC and considering trust as a factor in using a dataset.
Due to the small sample size of non-POC respondents, we conducted a Fisher's Exact test ($p=0.0207$, $OR=4.909$). 
The odds ratio indicates that the odds of considering trust in the decision to use a source for POC was 4.9 times higher than that of non-POC. We failed to find a significant relationship for the rest of our tests.

\subsubsection{Amount of data.}
We identified an association between the amount of data in a dataset contributing to the decision to use it and age group. 
A Chi-square test of independence revealed a significant relationship between the two factors ($p=0.0104$). 
Upon comparison of the expected and observed counts of the Chi-square test, we found that respondents under the age of 35 were more likely to consider amount of data in the decision to use a dataset.
The strength of this association was small but significant ($V=0.1662$). 
No other tests from this factor were significant.

\subsection{Experiences Finding Diverse Datasets (RQ2)}
To find the challenges and barriers that affect respondents’ ability to find diverse data, we asked questions regarding factors they believe impact their ability to find diverse datasets, the frequency with which they are able to find adequate data, and their experiences finding appropriate data.
We compared the responses to the independent variables: data use goals, age, and whether they identify as a POC. 
In this section, we will report the significant relationships from our tests.

\subsubsection{Factors that impact ability to find diverse sources.}
We identified a significant relationship between Q55 and age group. We conducted a Chi-square test of independence for the 3x3 contingency table and identified a significant relationship between factors selected and age ($p=0.8297$). 
We performed post-hoc testing to ensure the validity of results by calculating the adjusted residuals and correcting the significance level using a Bonferroni corrected alpha. 
From this analysis, we found that two factors have significant associations with age: \textbf{resources} (e.g, money, data sources) and \textbf{tooling} (e.g., language support). 
A comparison of expected and observed counts found that respondents under the age of 35 felt resources ($p=7.78e-07$, adjusted residual = $5.27$) impacted their ability to find diverse datasets more than tooling ($p=2.32e-5$,  adjusted residual = $-4.61$), whereas those over the age of 35 felt tooling impacted their ability more.
Cramer's V indicated a medium strength association ($V=0.3324$). 
Analysis of the other independent variables did not reveal any significant associations. 



\subsubsection{Difficulty finding trustworthy sources.}
We conducted a Mann-Whitney U Test to determine any associations between our independent variables and the ability to find adequate, representative datasets.
We identified a significant association between difficulty and identifying as a POC ($U=478.0$, $p=0.0005$).
This significant relationship encouraged us to evaluate the direction through a one-sided Mann-Whitney U test.
From this test, we determined that POC respondents found it more difficult to find trustworthy sources than non-POC respondents ($p=0.0002$). 
The mean difficulty score (on a Likert scale 1--5, Extremely Easy to Extremely Difficult) for POC was $3.1325$, while the mean for non-POC was $2.0$. 
We did not find significant relationships from the remainder of our tests.

\subsection{Finding relevant and representative datasets (RQ3)}
To better understand differences in experiences finding relevant and representative datasets, we compared the ability to find adequate and trustworthy sources for data with the first and most successful methods respondents use to find them.
In this section, we will elaborate on the significant results from our tests of association.

\subsubsection{Difficulty finding adequate data.}
Our analyses indicated a relationship between frequency not finding adequate data and beginning with a general web search ($U=5563$,$p=0.0061$).
Given the relationship, we performed a one-sided Mann-Whitney U Test to infer the direction.
Our results indicated that using a web search first was associated with less difficulty finding adequate data ($p=0.003$).  
The mean difficulty for beginning with a web search was $1.608$, while the mean difficulty for respondents who did not begin with a web search was $1.813$. 
The rest of our comparisons did not find any significant relationships. 


\section{Discussion}
Our findings thus far provide valuable insights for advancing our efforts and  others interested in the role of technologists and researchers in equitable data-centric innovation.\\

\noindent\textbf{The Role of Expertise in Data Seeking.}
Expertise plays a significant role in technologists' processes and strategies~\cite{latoza2020explicit}.
Prior research indicates significant relationships  between expertise in computing and age, emphasizing the heightened expertise found in older adults~\cite{arning2008development}.
Our findings suggest a relationship between age and the methods and considerations involved in seeking and using data for innovation, including the willingness to explore smaller datasets which studies have shown may be the case for datasets centered on historically marginalized groups~\cite{philip2013framework,warren2022increasing}.
We will use these insights to delve deeper into the role of expertise in data seeking behaviors, which can lead to broader, actionable insights for improving practice.\\

\noindent\textbf{Understanding Identity as a Factor in Data Seeking.}
Prior studies have provided insights into the role of identity and positionality in computing research and innovation~\cite{schwarz2005influence, scheuerman2024products, secules2021positionality}.
We found that the role of identity in innovation may extend to the intentionality behind and challenges with finding diverse datasets.
Our findings suggest racial disparities in data trust and accessibility, where POC reported greater difficulty finding trustworthy data sources. 
This points back to persistent systemic inequalities and underscores the importance of addressing issues of representation and bias in data collection and use.
In our efforts to better understand experiences and strategies involved in the usage of representative datasets, we will ensure that we are engaging with BIPOC technologists in research and development to better understand how we can more effectively support the use of representative datasets in practice.

\bibliography{references}

\section{Acknowledgments}
This work is supported by the National Science Foundation (NSF) under grant \#2224674 and \#2224675.


\end{document}